\documentclass[twocolumn,prb,superscriptaddress]{revtex4}
\usepackage{amssymb,amsfonts,amsmath}
\usepackage{graphicx}
\begin{document}

\title{Half-filled orbital and unconventional geometry of a common
  dopant in Si(001)} 
\author{K. Iwaya}
\affiliation{London Centre for Nanotechnology and
    Dept. of Physics and Astronomy, University College London, Gordon
    Street, London WC1H 0AH, UK}
\altaffiliation{Present address: Center for Emergent Matter Science, RIKEN, Saitama 351-0198, Japan}
\author{D. R. Bowler}
\affiliation{London Centre for Nanotechnology and
    Dept. of Physics and Astronomy, University College London, Gordon
    Street, London WC1H 0AH, UK}
\affiliation{TYC@UCL, University College London, Gower Street, London WC1E 6BT, UK}
\author{V. Br\'azdov\'a}
\affiliation{London Centre for Nanotechnology and
    Dept. of Physics and Astronomy, University College London, Gordon
    Street, London WC1H 0AH, UK}
\affiliation{TYC@UCL, University College London, Gower Street, London WC1E 6BT, UK}
\author{A. Ferreira da Silva}
\affiliation{Institute of Physics, Federal University of Bahia, 40210 340 Salvador, Bahia, Brazil}
\author{Ch. Renner}
\affiliation{Department of Condensed Matter Physics, University of Geneva, Geneva, CH-1211, Switzerland}
\author{W. Wu}
\affiliation{London Centre for Nanotechnology and
    Dept. of Physics and Astronomy, University College London, Gordon
    Street, London WC1H 0AH, UK}
\affiliation{TYC@UCL, University College London, Gower Street, London WC1E 6BT, UK}
\author{A.J. Fisher}
\affiliation{London Centre for Nanotechnology and
    Dept. of Physics and Astronomy, University College London, Gordon
    Street, London WC1H 0AH, UK}
\affiliation{TYC@UCL, University College London, Gower Street, London WC1E 6BT, UK}
\author{A. M. Stoneham}
\affiliation{London Centre for Nanotechnology and
    Dept. of Physics and Astronomy, University College London, Gordon
    Street, London WC1H 0AH, UK}
\affiliation{TYC@UCL, University College London, Gower Street, London WC1E 6BT, UK}
\author{G. Aeppli}
\affiliation{London Centre for Nanotechnology and
    Dept. of Physics and Astronomy, University College London, Gordon
    Street, London WC1H 0AH, UK}



\begin{abstract}
The determining factor of the bulk properties of doped Si is the column rather than the row in the periodic table from which the dopants are drawn. It is unknown whether the basic properties of dopants at surfaces and interfaces, steadily growing in importance as microelectronic devices shrink, are also solely governed by their column of origin. The common light impurity P replaces individual Si atoms and maintains the integrity of the dimer superstructure of the Si(001) surface, but loses its valence electrons to surface states. Here we report that isolated heavy dopants are entirely different: Bi atoms form pairs with Si vacancies, retain their electrons and have highly localized, half-filled orbitals. 
\end{abstract}
\maketitle

\section{Introduction}
\label{sec:introduction}




\begin{figure}
  \includegraphics[width=0.4\textwidth]{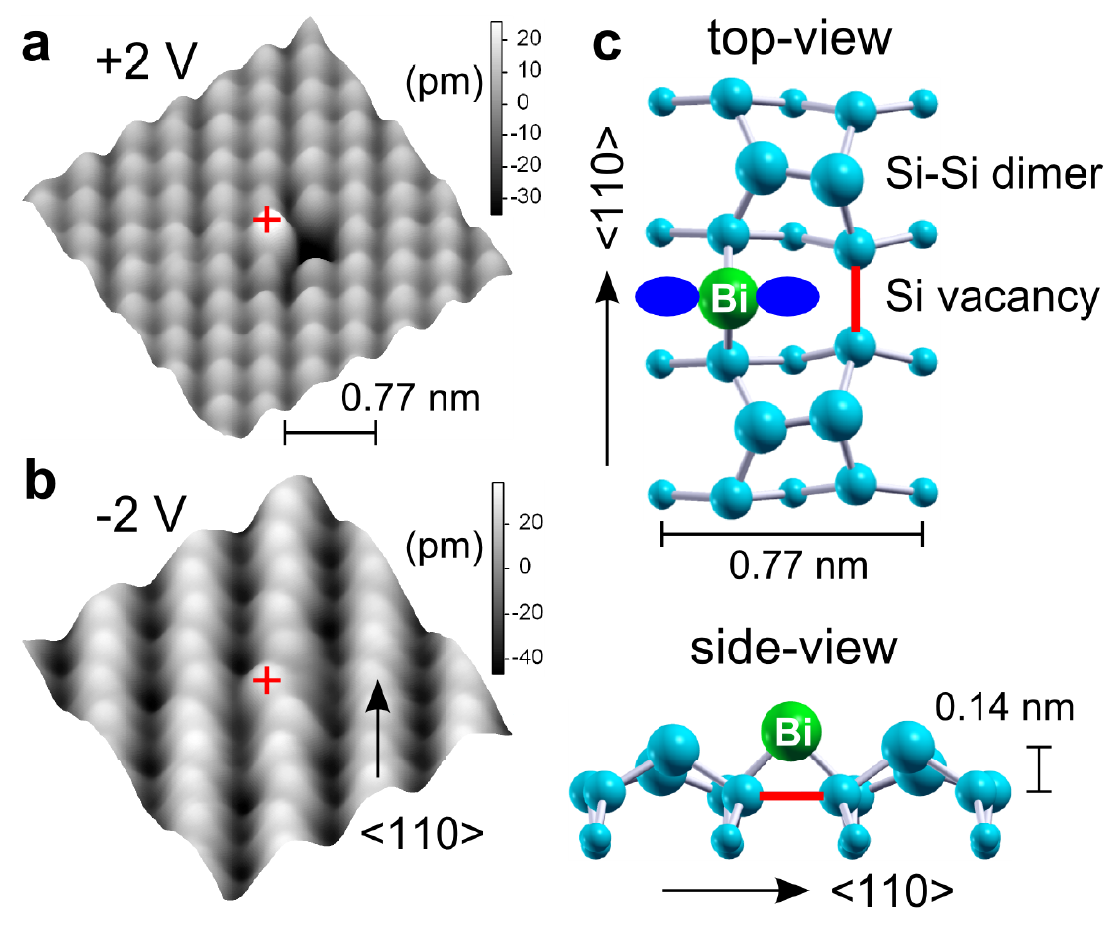}
  \caption{Structure of Bi-Si vacancy complex in the Si(001) surface.
  (a, b) Bias-dependent topographic images of Bi donor in the Si(001)
  surface at sample-bias voltages of V$_{s}$ = +2 V (empty state) and -2 V
  (filled state), respectively. The tunneling current was I$_{t}$ $\sim$ 100
  pA. Bi sites are indicated by markers. The color scale is in units
  of pm. The horizontal scale is indicated in (a) by the spacing
  between Si-Si dimer rows ($\sim$0.77 nm). (c) Schematic structure of
  Bi–Si vacancy complex in the Si(001) surface (top-, side-view). The
  neutral Bi atom has a half-filled p orbital perpendicular to the Si
  dimer row, and at the Si vacancy site the second-layer Si atoms bond
  across the vacancy as shown by a dotted line.}\label{fig:structure}
\end{figure}

Semiconductors such as silicon (Si) derive remarkable utility from the
ease with which they are converted from insulating to conducting behaviour via
substitution of dopants from neighbouring columns in the periodic
table. Isolated phosphorus (P) and bismuth (Bi) atoms, both with one
more valence electron than Si, substitute for Si atoms, and behave
approximately as hydrogenic atoms with radii and binding energies
described by the Bohr theory where the electron mass and dielectric
constant are those provided by the diamond lattice of Si rather than
the vacuum. 

Dopants in the bulk are well understood, but their
properties at surfaces are poorly understood; for instance, we have recently shown
that heavy dopants at Si(111) can induce ambipolar charge
states\cite{Studer:2012yq}.  
There is growing interest in heavier dopants of conventional
semiconductors because of their more tightly bound outer electrons,
which increases operating temperatures for applications ranging from
THz impurity lasers to quantum computers exploiting their spin\cite{Morley:2010gf}. We
have accordingly chosen the heaviest column V element for the current
study. 
Here we report that isolated heavy
dopants in the Si(001) surface are entirely different to light
dopants: Bi atoms form pairs with Si vacancies, retain their electrons
and have highly localized, half-filled orbitals. 
Thus, we finally have
a dopant of silicon with a half-filled orbital at the surface,
enabling both new applications in spintronics, as well as Si
surface-based quantum many-body physics, hitherto confined to the more
complex Si(111) surface.

\section{Methods}
\label{sec:methods}

Our samples are semiconductor grade p-type Si(001) wafers (with
boron [B] doping of $\sim$ 10$^{15}$ cm$^{-3}$) implanted with Bi$^+$ ions (Si:Bi). The
Bi donor concentration estimated from transport measurements is
$\sim$ 2.2$\times$ 10$^{19}$ cm$^{-3}$, which is near the
metal-nonmetal transition\cite{Abramof:1997gf}.  

Scanning tunnelling microscopes (STMs), where a metallic tip is
positioned to sub-Angstrom precision using piezoelectric drives and
the tunnelling current between tip and sample is directly measured
and/or used to control the drives, are uniquely suited to determine
the electronic structure of surfaces. Here we employ an Omicron
low-temperature (LT)-STM, at 5 K in ultrahigh vacuum (UHV) with a
base pressure of $\sim 4 \times 10^{-9}$ Pa. For the Si(001) surface, strong
tip-surface interactions can induce a complicated, unstable
structural change between c(4$\times$2) and p(2$\times$2) reconstruction at low
temperature\cite{Yoshida:2004oz,Sagisaka:2005xr}. To avoid this, we ensure a relatively large tip-sample
separation during measurements by using small tunnelling currents,
typically 100 pA for topographical imaging. 

The surface was prepared in UHV by a standard flashing (direct current
heating) method at $\sim$1100$^\circ$C for 10 seconds. The pressure was kept
below 1$\times$ 10$^{-8}$ Pa during the flashing to prepare a clean surface. When
searching for features resulting from Bi dopants, we must first
eliminate other features. The electronic structure of standard defects
in a clean Si(001) surface have already been studied by STM\cite{Hata:2000fk,Hamers:1989lr,Brown:2002qy}, and
so such defects can be identified and eliminated from our study by
bias-voltage-dependent imaging. The native B dopant might also be
observed; however, it has a much lower density than the ion-implanted
Bi impurities, and bias-dependent STM images of a single B dopant in
Si(001) surface have already been reported\cite{Zhang:1996uq}, so also allowing B
impurities to be identified, if present, and excluded from the current
study. 

The electronic structure calculations used density functional theory (DFT) within the PBE generalized
gradient approximation (GGA)\cite{Perdew:1996kx}. The system was modelled with a slab at least six
layers deep with the last two layers fixed in bulk-like positions and
the bottom layer terminated in hydrogen. The Brillouin zone was
sampled with the same density for all cells at a level which gave meV
convergence in total energies.  We used two different basis sets and
codes: plane waves and gaussians.


The plane-wave basis calculations as implemented in the VASP
code\cite{kresse1996b} used a cutoff energy of 200 eV,
which converged energy differences to meV accuracy.  We used ultrasoft
pseudopotentials\cite{Vanderbilt:1990fj} and spin-polarized density
functional theory, with the PBE GGA functional\cite{Perdew:1996kx} for
the exchange-correlation terms. After performing spin-polarized
calculations we also tested the effect of including spin-orbit
interactions, which made only a small difference to the electronic
structure. For calculations on a single dopant, we chose the cell
sizes to vary the spacing between its periodic images. We used cells
with between 1 and 3 dimer rows, giving lateral sizes of 0.77 nm, 1.54
nm and 2.30 nm, with four to eight dimers in the cell, giving lengths
of 1.54 nm and 3.07 nm. The basic density of k-points was set by the
1.54 nm $\times$ 1.54 nm cell, which we sampled with a mesh of $4
\times 4 \times 1$ k-points; other cells had the number of k-points
varied to achieve the same density, and convergence checks were made
on all cells. Structural relaxations were carried out until all
components of all forces were less than 0.25 eV/nm.  

Exact exchange is relatively inefficient to implement in plane-wave
DFT because the exchange kernel is non-local; we therefore use the
CRSYTAL06 code\cite{Dovesi:2006zt}, which has a Gaussian basis set,
using the same atomic coordinates as were found by the VASP
calculations.  The basis sets used were 88-31G$^\star$ for Si and Bi
and 31G for the terminating H atoms on the reverse of the slab, with
an all-electron treatment of the Si atoms and an effective core
potential\cite{Watd:1985xw} for the Bi atoms.  The k-point sampling
was a (2$\times$1$\times$1) mesh, giving the same density as in the
VASP calculation.

\section{Results}
\label{sec:results}

\subsection{Atomic Structure}
\label{sec:atomic-structure}

The Si(001) surface shows characteristic $\langle 110\rangle$ rows of
dimerized Si atoms\cite{Tromp:1985dk}. Dimers along the row are
separated by $a_0/\sqrt{2}=0.38$nm, where $a_0$ is the cubic lattice
constant of Si.  Tunnelling images of the Bi-implanted samples after
in-situ surface preparation contained four easily distinguished
features attributable to Bi donors. The bismuth-vacancy feature we
describe in this work, referred to as Bi-V, is located at the surface;
we have found other features which are clearly sub-surface features
and will be discussed in a future publication; these sub-surface
donors are considerably less common than the surface Bi-V pairs. Of
our large inventory of 500 observed features, $\sim$84 \% are Bi-V
pairs and are located at the surface. Figs. \ref{fig:structure}a and
\ref{fig:structure}b display empty- and filled-state topographs of a
single Bi-V defect and how it fits within the dimer rows. For filled
states (Fig. \ref{fig:structure}b; a sample-bias-voltage V$_S$ of -2
V) the Bi-V feature is similar to the surrounding surface, though a
little brighter. For empty states (Fig. \ref{fig:structure}a; V$_S$ =
+2 V), it appears as a bright protrusion neighbouring a depression
surrounded by the characteristic dimer $\pi^\star$-bonding states;
this is a very unusual structure in the Si(001) surface. The apparent
flatness of the surface surrounding the Bi-V feature indicates that
the Bi is not charged---strong brightening is seen around charged
dopants in Si(001)\cite{Radny:2006fe}, due to band bending.
Considering the topographic structure, it is extremely likely that the
Bi atom substitutes for a Si atom in the surface, as illustrated in
Fig. \ref{fig:structure}c. 

\begin{figure}
  \centering
  \includegraphics[width=0.5\textwidth]{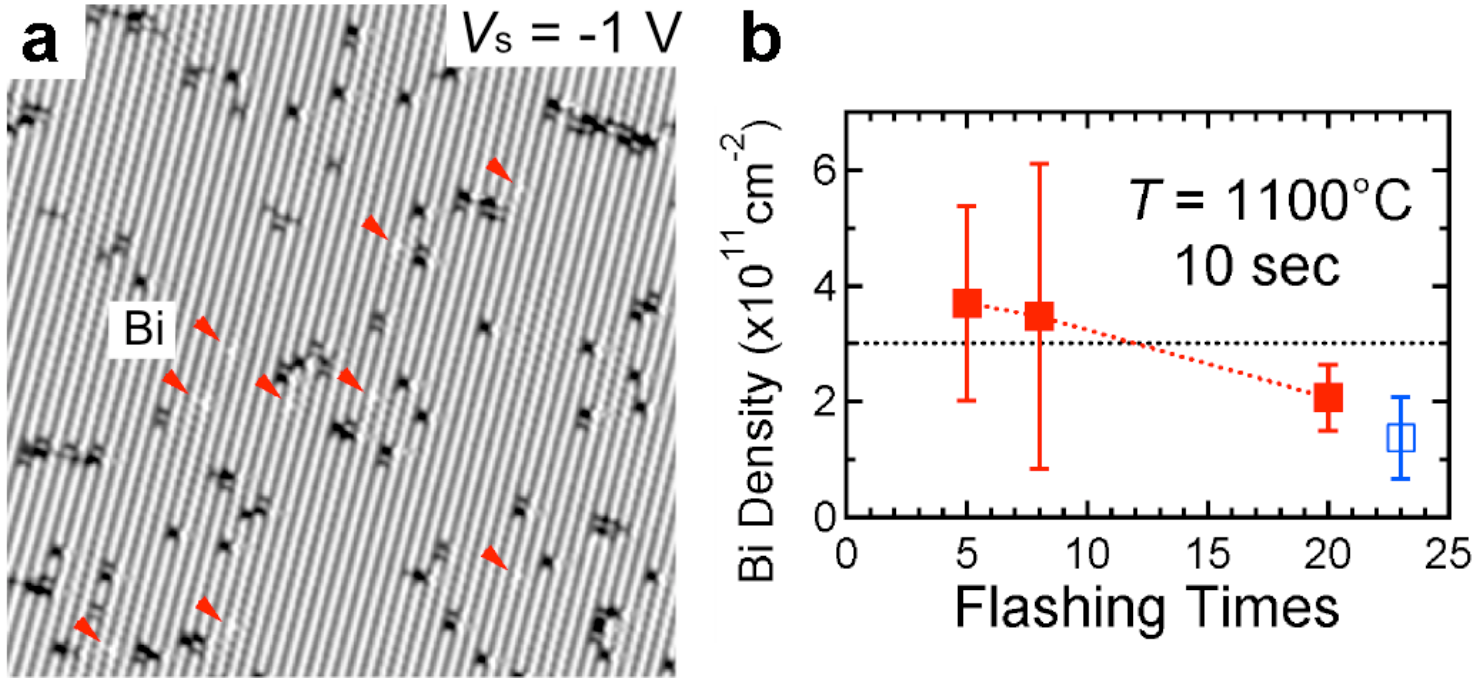}
  \caption{(a) Typical topographic image of Si(001):Bi at Vs = -1
    V. Scan size is 40 nm $\times$ 40 nm. Bi donors are indicated by
    arrows. (b) Flashing effect on the density of Bi dopants in STM
    image. The flashing was performed at 1100$^\circ$C for 10 sec. The
    sample was exposed to air after 20th flashing intentionally, and
    flashed again in UHV condition (blue square). For comparison, the
    horizontal dashed line represents the Bi concentration estimated
    from the bulk density ($\sim 3 \times 10^{11}$ cm$^{-2}$). } 
  \label{fig:SI1}
\end{figure}

As a further check on the identity of the new features we investigated the stability of the Bi-V
structure and its population density by repeated sample flashing. The density
is estimated from topographic images (Fig.~\ref{fig:SI1}a)  to
be $\sim 2 \times 10^{11}$ cm$^{-2}$ after repetitive surface flashing
(Fig.~\ref{fig:SI1}b). If we assume that Bi atoms in the first two
layers are visible in STM, and have the same density as produced by
the initial ion implantation/annealing protocols, then we expect to
see ($2.2 \times 10^{19}$ cm$^{-3}$ ) $\times$ (0.14 nm) $\sim 3
\times 10^{11}$ cm$^{-2}$ (as Si layer spacing along (001) is ~0.14
nm). This is in excellent agreement with the defect density observed
by STM, strengthening our identification of these features as Bi
donors.  Moreover, the Bi-V structure is always found, with a
population density within a factor of two of that associated with the
bulk dopant density. Even after the sample was exposed to air
intentionally and heated again, the same defects are seen, although
with a barely significant trend towards lower density with repeated
surface treatments.

\begin{figure}
  \includegraphics[width=0.48\textwidth]{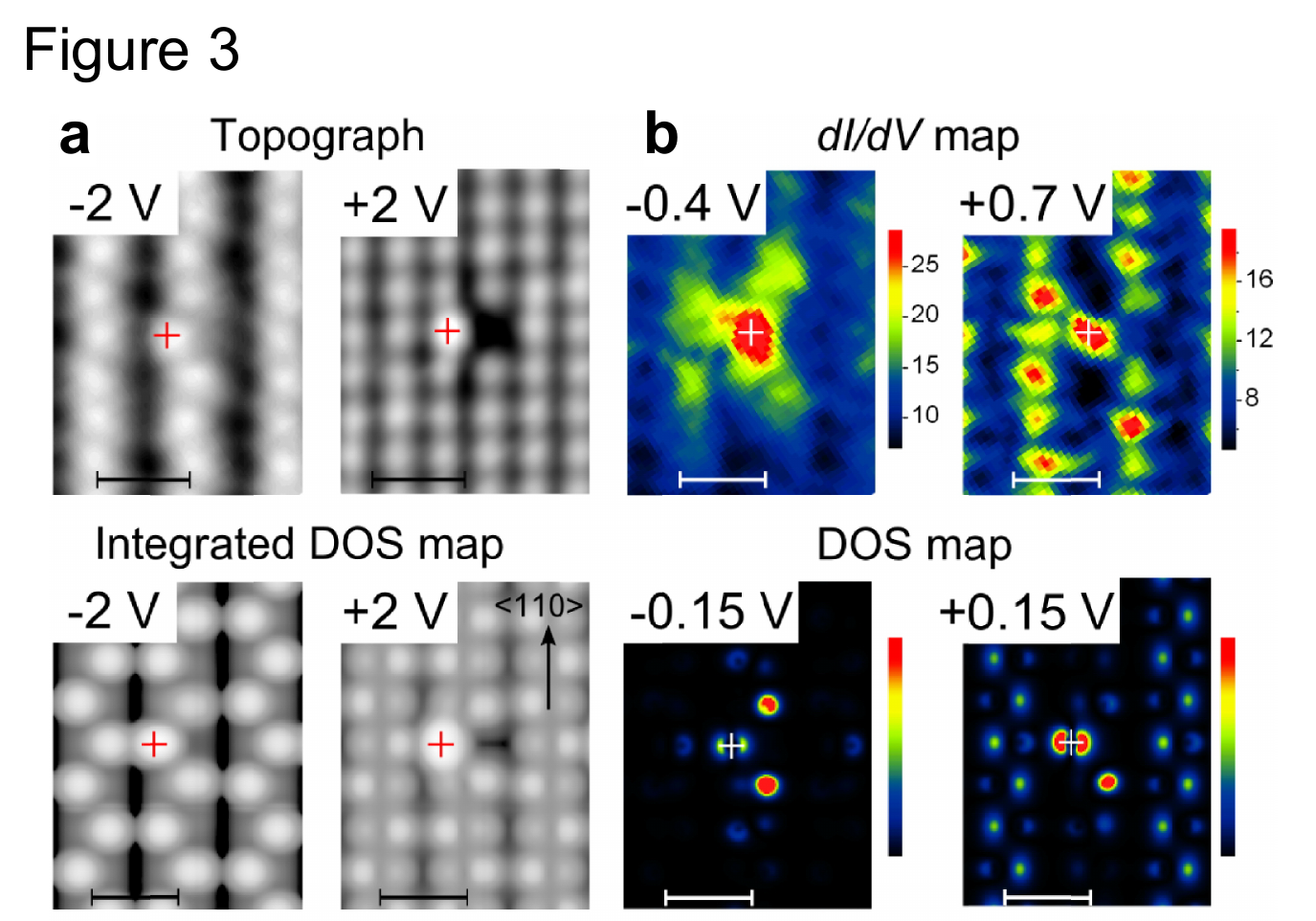}
  \caption{Comparison between STM/STS and spin-polarized DFT.
(a) STM images (top) and simulated STM (bottom) of the Bi
donor (plus sign).  STM images were taken at exactly the same
location; simulated images used isosurface values of 0.1
electron/nm$^{3}$. (b) dI/dV maps (top) and
simulated DOS maps around Bi-Si vacancy complex (bottom). Feedback
set-point was $\sim$ 100 pA, Vs = -2 V. The spacing between
Si-Si dimer rows ($\sim$0.77 nm) is shown in all images. The units of color
scale in the dI/dV map and DOS map are pS and electron/nm$^{3}$.}\label{fig:comparisona} 
\end{figure}

To identify the structure of the feature and further characterize it,
we have performed total energy spin-polarized density functional
theory (DFT) calculations.   As the structure clearly involves a
vacancy and the prepared surface contains a certain number of
vacancies, we have considered structures involving a vacancy; the
energetics of structures in the absence of vacancies have also been
considered and are given below.  Various candidate structures were
considered: a missing dimer vacancy (1DV) and a mixed Bi-Si dimer with
different charge states, by
analogy with the P-Si hetero-dimer system\cite{Reusch:2007la,Reusch:2006pb,Radny:2006fe}; a Bi adatom on the
Si(001) surface with a 1DV; a Bi atom symmetrically replacing an entire Si dimer
in the surface; and a Bi-vacancy pair replacing a Si dimer in the
surface, again with different charge states. Kinetic effects during sample
preparation produce an abundance of non-equilibrium
structures\cite{Itoh:2000rr}. Accordingly, the main criterion used to judge the fitness
of the structure was the agreement of the appearance between simulated
and real STM images, rather than total energy from DFT; the stability
of different features is discussed further below.  

\begin{table}
\caption{Energies calculated with plane wave and gaussian basis set
  spin-polarized DFT of possible structures relative to Bi atom placed 
  symmetrically in missing dimer vacancy (Symmetric Bi).  For
  Symmetric Bi and Bi-V the energies relative to symmetric Bi
  calculated with hybrid DFT are also shown.  Structures are illustrated in Fig.\protect~\ref{fig:SI2}}
\label{tab:energies}
\begin{tabular}{ccc}
Structure & DFT (eV) & Hybrid DFT (B3LYP) (eV) \\
Symmetric Bi & 0.00 & 0.00 \\
Bi-V & 0.39 & -0.04 \\
Bi adatom + 1DV & 0.67 & -- \\
Bi-Si HD + 1DV & 0.79 & -- 
\end{tabular}
\end{table}
 
\begin{figure}
  \centering
  \includegraphics[width=0.5\textwidth]{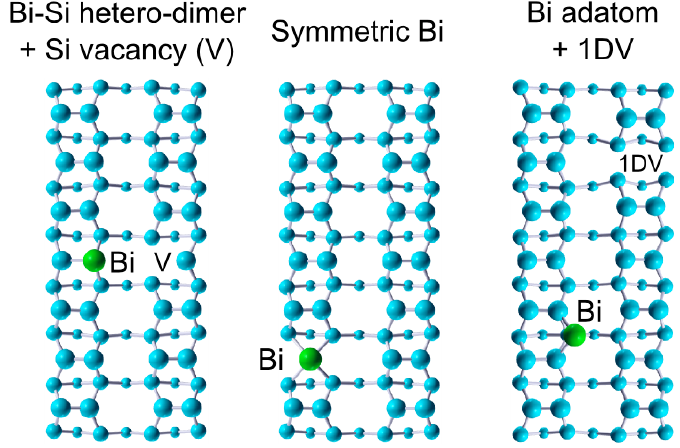}
  \caption{Schematic structures of arrangements of Bi and vacancies on Si(001). ``1DV'' is the well-known single Si dimer vacancy, and ``V'' is a single Si vacancy. }
  \label{fig:SI2}
\end{figure}

The only structure which accounts for the STM images is the neutral
surface Bi-vacancy pair (Bi-V). Fig. \ref{fig:structure}c is the associated schematic,
while Fig. \ref{fig:comparisona}a allows comparison of STM topographic images and DFT
calculations. In this model, one atom of a surface Si dimer is
replaced by a two-fold coordinated Bi atom, while the other Si atom is
removed, and the second-layer Si atoms rebond across the vacancy as
indicated by a red line in Fig.~\ref{fig:structure}c. The resulting Bi atom has a neutral
configuration with two 6$s$ and three 6$p$ electrons, unhybridized, in
contrast to that in bulk Si where the donor ionizes leaving four
electrons in $sp^3$ hybrid orbitals. This structure, and other unusual Bi
structures at the Si(001) surface such as Bi nanolines\cite{Owen:2002fy}, can be
explained by considering the large energy difference between the 6$s$
and 6$p$ Bi valence electrons\cite{Bowler:2000uf}: the hybridization energy required to form
$sp^3$ orbitals, giving the tetrahedral bonding characteristic of Si, is
very high, and structures allowing $p^3$ bonding (as in native Bi bulk)
will be more stable. Density of states (DOS) plots for the Bi-V
structure show that the Bi 6$s$ orbital can almost be considered as a
part of the core, with only the 6p electrons forming valence
bonds. Two of the three 6$p$ orbitals form bonds with second-layer Si
atoms, and the final 6$p$ orbital remains unbonded and half-filled,
aligned in the plane of the surface and perpendicular to the Si-Si
dimer row direction as shown in Fig. \ref{fig:structure}c. The DOS of the Bi thus
originates mainly from 6$p$ orbitals, and the contributions from $s$ and $d$
orbitals are small in the energy range studied here.

Of the structures which we have modelled covering the possible atomic
arrangements, only the Bi-V structure matches the bias-dependent STM
images and spectroscopy.  Of all structures involving vacancies,
DFT finds the symmetric Bi structure to be most stable, with the Bi-V
structure the next most stable; however, a hybrid DFT calculation
finds that this ordering is reversed, with the Bi-V structure most
stable (see below). A possible formation mechanism for the
Bi-V structures at the surface is the diffusion of Bi atoms through
the bulk in association with Si vacancies\cite{Fahey:1989oq}. Further DFT calculations,
also detailed below, indeed show that
the Bi and Si vacancy will remain associated at the surface, once paired.

We give the total energies of various structures in Table~\ref{tab:energies}, from
both DFT GGA and hybrid DFT calculations, as DFT will underestimate the
energy of localised states thus also underestimating the stability of
features such as the Bi-V.  The structures modelled are illustrated
in Fig.~\ref{fig:SI2}.  The energies are all calculated for unit
cells with the same number of atoms (hence we have included a surface
vacancy in the calculations where Bi is not associated with a vacancy).

It is important to note various points about the energies. First,
once a vacancy is present at the surface, the Bi-V or Symmetric Bi
structures are most stable, as shown in
Table~\ref{tab:energies}. Second, given the non-equilibrium
preparation technique, structures which are energetically unfavourable
can form because kinetic pathways---derived from the highly
non-equilibrium ion implantation process and subsequent incomplete
thermal annealing steps---favour their formation. Third, while the
symmetric Bi feature has a lower energy, it does not match the feature
observed in STM. Fourth, calculations with hybrid functionals
indicate that, for the B3LYP functional, the Bi-V
structure is more stable than the symmetric Bi\footnote{The precise
  ordering of the structures is dependent on the effective core
  potential and the basis set used as well as on the
  exchange-correlation potential.}.  Finally, significant numbers of
energetically unfavourable structures (e.g. defects, steps) are found
in clean, well-prepared Si(001) surfaces in all experiments. 

\subsection{Electronic Structure}
\label{sec:electronic-structure}

\begin{figure}
  \includegraphics[width=0.48\textwidth]{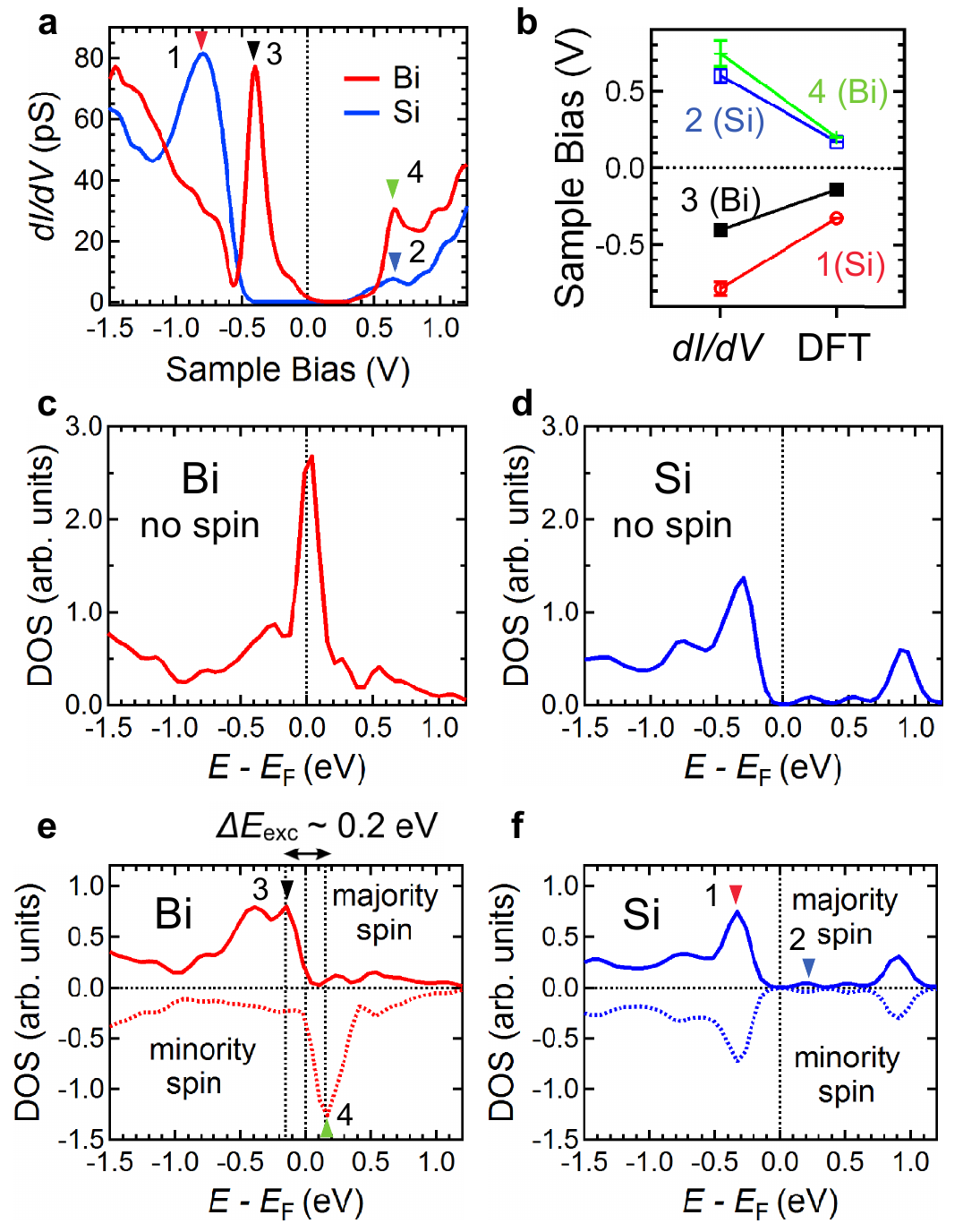}
  \caption{Comparison between STS and both non-spin-polarized and spin-polarized
    DFT simulated density of states (DOS) of Bi - Si vacancy complex.
(a) Differential conductance dI/dV at Bi donor and Si dimer far
from Bi.  Characteristic peaks are indicated by arrows and numbers. 
(b) Comparison of peak bias-voltages between dI/dV and spin-polarized calculated DOS. 
(c) DOS of Bi atom calculated by non-spin-polarized DFT. 
(d) Simulated DOS of Si far from Bi by non-spin-polarized DFT. 
(e) Spin-polarized DOS of Bi atom embedded in the supercell shown in
Fig. \ref{fig:structure}c. The calculated exchange energy is $\sim$
0.2 eV.   Peaks are indicated with the same colors and numbers as for dI/dV.  
(f) Spin-polarized DOS of Si far away from Bi.}\label{fig:compareDOS} 
\end{figure}

To understand the electronic structure of the Bi-V feature, we
performed STS measurements and compared them to non-spin-polarized and spin-polarized DFT DOS
calculations. Fig. \ref{fig:compareDOS}a shows the differential conductance (dI/dV)
spectrum collected with the STM, both at the Bi site and over a Si
dimer far from the Bi.  The simulated local density of
Kohn-Sham states (LDOS) around the Bi dopant and the upper Si atom of
a buckled Si-Si dimer far from the Bi site are shown for spin
non-spin-polarized calculations in Fig. \ref{fig:compareDOS}c and d and for
spin-polarized calculations in Fig. \ref{fig:compareDOS}e and f. Note
that the non-spin-polarized LDOS are very different to the experiment, while
the spin-polarized agree well (discussed in more detail below).  Our key new result is a
double-peaked Bi-induced resonance in the Si dimer gap, visible in
both the spin-polarized DFT and STM.

We compare experiment with theory first for clean Si. The dI/dV
spectrum taken at the Si dimer shows characteristic peaks at $\sim$ -0.8,
+0.6 V (Fig. \ref{fig:compareDOS}a), which are in agreement with previously reported
results\cite{Yoshida:2004oz,Sagisaka:2005xr}. Similar features representing $\pi$ (-0.4 V) and $\pi^\star$ (+0.2 V)
states of the Si-Si dimer are seen in the simulated LDOS plot
(Fig. \ref{fig:compareDOS}d and f). It is well-known that DFT significantly underestimates band
gaps, though it will give reasonable state densities and charge
density distributions once appropriate energy shifts are made\cite{Martin:2004wd}. This
will contribute strongly to the energy differences between experiment
and theory: $\sim$ 0.4 V in both
empty and filled states.  Another factor
is likely to be tip-induced band bending\cite{Feenstra:2002rw}. 

 
The experimental dI/dV spectrum of Bi in Fig. \ref{fig:compareDOS}a shows a
sharp peak at $\sim$ -0.4 V and a small peak structure at $\sim$ +0.7 V. The
-0.4 V peak can be attributed to the sharp peak at -0.15 V in the
simulated spin-polarized DOS (Fig. \ref{fig:compareDOS}e), associated with the Bi p
state. Similarly, the peak at +0.7 V in the dI/dV spectrum can be
matched to the +0.15 V peak in the calculated spin-polarized DOS. The extra peaks
seen in the calculated DOS are caused by slight mixing with
neighbouring Si atoms and are due to the method of projecting the DOS onto
individual atoms. Fig. \ref{fig:compareDOS}b  summarizes the
comparison of peak bias-voltages found for dI/dV and calculated DOS.

Fig. \ref{fig:comparisona}b shows dI/dV maps around the Bi-V
complex at two characteristic peak energies, -0.4 V and +0.7 V. The
filled state map at -0.4 V reveals the spatial structure of the Bi
donor level. The DOS around Bi becomes larger reflecting the sharp Bi
p state peak, and a substantial hybridization with neighbouring Si
atoms can also be observed. In the empty states at +0.7 V, the Bi site
appears as a localized protrusion. To compare the experimental results
with DFT calculations, including the energy shifts due to DFT errors
and band bending, we have shifted the simulated DOS map in accord with
the shifts of the key spectral features in Fig. \ref{fig:compareDOS}a. Thus, the
simulated DOS map at -0.15 V (corresponding to -0.4 V in dI/dV map)
can account for the spatial structure of the Bi donor level observed
in the dI/dV map. Also, the localized bright spot in the dI/dV map at
+0.7 V is clearly reproduced at +0.15 V in the DFT calculation. The
electronic structure of a Bi-Si heterodimer is shown in
Appendix~\ref{sec:comp-heter}, and confirms that the Bi-V feature is
not a heterodimer.


\subsection{Spin and Magnetic Structure}
\label{sec:spin-magn-struct}

 Fig. \ref{fig:compareDOS}c--f present non-spin-polarized and polarized densities of
 states (DOS) for Bi and Si atoms. Without polarization, which is
 required unless the Bi is ionised, the calculated DOS of Bi is
 inconsistent with the measured dI/dV spectra, since a single peak at
 E$_{F}$ is predicted (Fig. \ref{fig:compareDOS}c). Taking spin into account, the majority
 band associated with the unbonded Bi p-state is full and the minority
 band empty (Fig. \ref{fig:compareDOS}e, left), implying full polarization, with a moment of 1
 $\mu_B$/Bi, and the theory bears much closer resemblance to the data in
 Fig. \ref{fig:compareDOS}a. The associated spin density is dominated by the Bi atom and
 one of the ‘up’ Si atoms neighbouring the Si vacancy on the opposite
 side of the dimer row, but shows some delocalization along the row.
 There is no significant difference between polarized and non-spin-polarized
 results (Fig. \ref{fig:compareDOS}d and f) for the DOS at the Si sites far from the
 defect. Thus, the peaks seen in the dI/dV spectra for Bi-V must be
 due to a spin-polarized Bi state.  
 The splitting between the occupied and unoccupied states at a single
 spin-polarized defect is approximately 0.2 eV, a large value
 supported by our experiments. 

\begin{figure}
  \centering
  \includegraphics[width=0.5\textwidth]{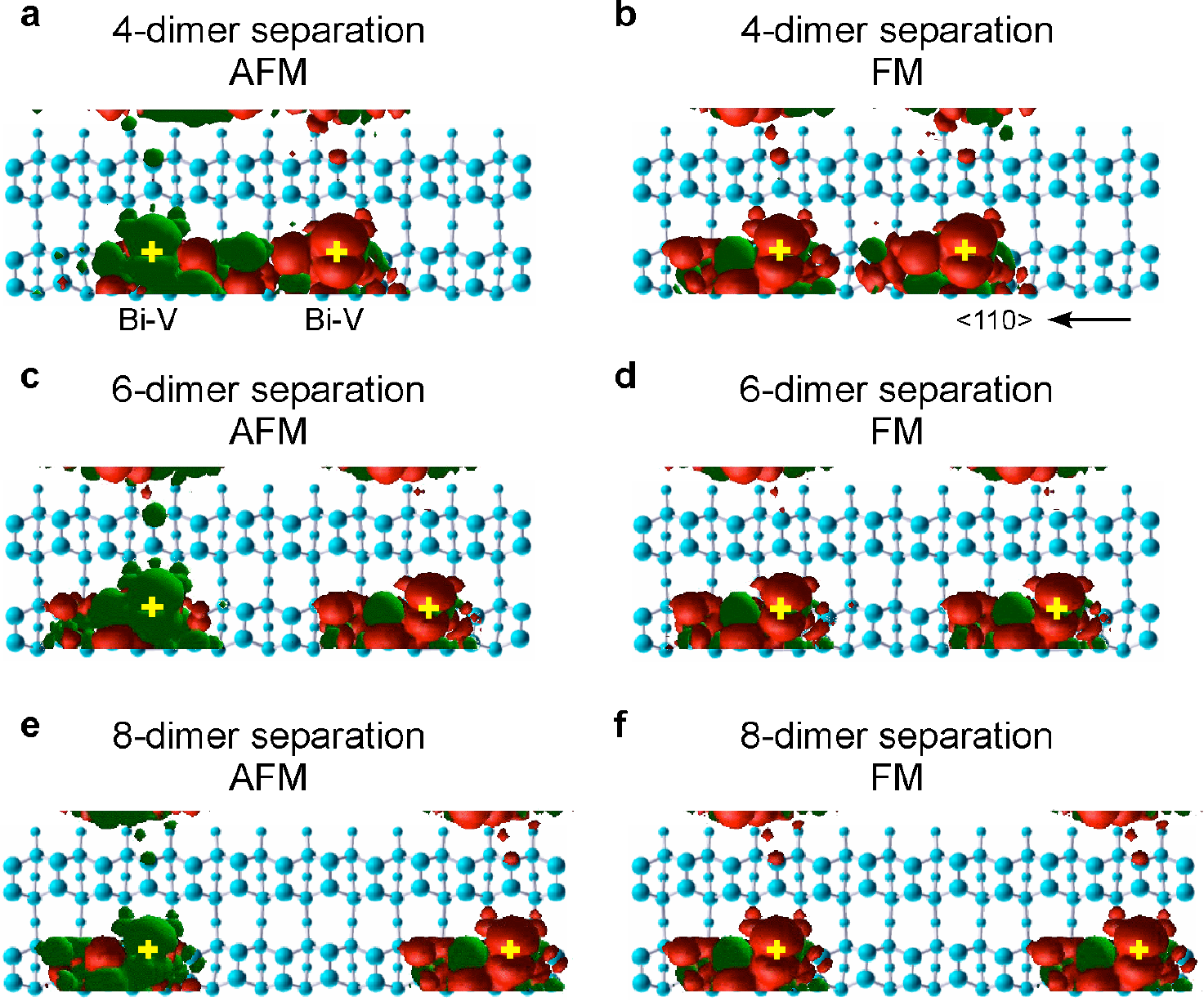}
  \caption{Contours of constant spin density for broken-symmetry AFM configurations (left column) and FM configurations (right column) of Bi-V pairs on Si(001), computed using hybrid DFT. The dimer rows run from left to right; red and blue iso-surfaces show positive and negative spin densities respectively. (a)-(b); spacing of four dimers. (c)-(d); spacing of six dimers . (e)-(f); spacing of eight dimers.  }
  \label{fig:SI4}
\end{figure}

Though we are unable at present to show experimental data in
support of this, we now turn to calculations of the
energetics of magnetic ordering between pairs of Bi-V features. 
We compare the ferromagnetic and broken-symmetry
antiferromagnetic total energies in order to obtain the couplings of
the Bi-V pairs.  Because of the p$^3$ bonding and the size of the
atoms, the Bi atoms will always be found substituting for the `up'
position of a buckled dimer, with the vacancy at the `down' site.  We
have computed only pairs where the Bi atoms are located on the same
side of the dimer row; because of the `up-down' tilt sequence found at
low temperatures in the Si(001)-c(4$\times$2) structure, this implies
a spacing of an even number of dimer separations (and hence an odd
number of dimers separating the two defect sites). 

 DFT relies on
 a finite supercell, with two rows of eight dimers each, to obtain the
 DOS.  It is natural to ask to what extent the splitting is derived
 from the ferromagnetism implied by the periodic boundary
 conditions. Accordingly, we have also considered the electronic
 structure of arrays of pairs of the Bi-V feature in the same
 supercell with different separations, shown in Fig.~\ref{fig:SI4}. For these
 pairs we can estimate the exchange interaction between neighbouring
 Bi atoms by calculating the energy difference between ferromagnetic
 (FM) and antiferromagnetic (AFM) arrangements of the spins localized
 on the Bi atoms. Since it
 is known that conventional DFT tends to delocalize electronic states
 artificially\cite{Zunger:2010qe}, and hence over-estimate such magnetic interactions,
 we have also made this comparison using hybrid DFT.

 The calculations considered interactions when Bi atoms are in the
 same row, using two rows of sixteen dimers each, with the size is
 constrained by the limits of DFT; all results are shown in
 Fig. \ref{fig:SI4}.  We consider Bi atoms on the same side of the
 row, with spacings of 4, 6 or 8 dimers, or 1.54, 2.30 and 3.07 nm.
 The FM/AFM energy differences derived from hybrid DFT are: -0.03eV
 for 1.54 nm (4 dimers); they also favour ferromagnetism but have
 magnitudes below 0.001 eV for 2.30 and 3.07 nm (6 and 8 dimers). This
 difference is significant for the 4-dimer spacing, but falls below
 the accuracy of our calculations for 6 and 8 dimers.  The
 calculations indicate that the interactions are ferromagnetic, in
 contrast to the antiferromagnetic interaction of around +0.06 eV
 estimated for Bi at the same separation in bulk Si\cite{Wu:2008kl},
 with considerable overlap of the associated Rydberg atoms. The
 difference is a result of the bonding of the Bi and hence its
 electronic structure.

There are two significant difficulties in analysing such magnetic
couplings between defect pairs using DFT: first, conventional DFT
tends to delocalize artificially localized spin densities, and hence
over-estimate magnetic couplings\cite{Zunger:2010qe}.  Second, even in
pairs on the insulating side of a Mott transition, the Kohn-Sham
orbitals form delocalized molecular states over both members of the
pair and the bonding state is doubly occupied in the case of
anti-aligned spins (anti-ferromagnetic pair), whereas for aligned
spins both bonding and anti-bonding states are singly occupied.  This
artificially raises the energy of AFM relative to FM structures.  We
solve the first problem by using the B3LYP hybrid
functional\cite{Becke:1993il}, which adds an empirical mixture of
exact exchange into gradient-corrected DFT; this is found to give a
better balance between localization and delocalization and to give
magnetic couplings in better agreement with experiment in a range of
structures\cite{Zunger:2010qe,Becke:1993il}.  We deal with the second
problem by using a broken-symmetry approach\cite{Noodleman:1981ee}, in
which the Kohn-Sham states are no longer required to transform
according to irreducible representations of the structural symmetry
group but are rotated so as to localize them on the magnetic centres
and enable a fair comparison of the FM and AFM configurations. 

The resulting spin densities are shown in Fig.~\ref{fig:SI4}.  The
strong two-lobed feature corresponding to the unbonded Bi p-state is
clearly visible in all cases, as is the delocalization of the spin
density onto the dangling bond for a neighbouring Si up-atom. In most
cases this is on the opposite side of the dimer row, but for the
smallest separation considered (4 dimer spacings) the strain in the
surface has caused the neighbouring dimer of the right-hand defect
shown to reverse its tilt, and accordingly the spin density spills
onto the neighbouring Si on the same side of the dimer row
instead. The origin of the magnetic coupling in the delocalization of
the spin density along the dimer row is clearly visible in the 4-dimer
spacing case, and can also be seen in lower-density isosurfaces (not
shown) for the larger separations.

The localisation of the Bi state on the surface is significantly
greater than in bulk, but it can couple to the delocalised $\pi^\star$
state on the surface along dimer rows; in the limit of no dimer tilt,
where the surface bands are half-filled and therefore metallic, the
resulting RKKY interaction will give alternating ferromagnetic and
antiferromagnetic interactions along a row.

\section{Conclusion}
\label{sec:conclusion}

\begin{figure}
  \includegraphics[width=0.3\textwidth]{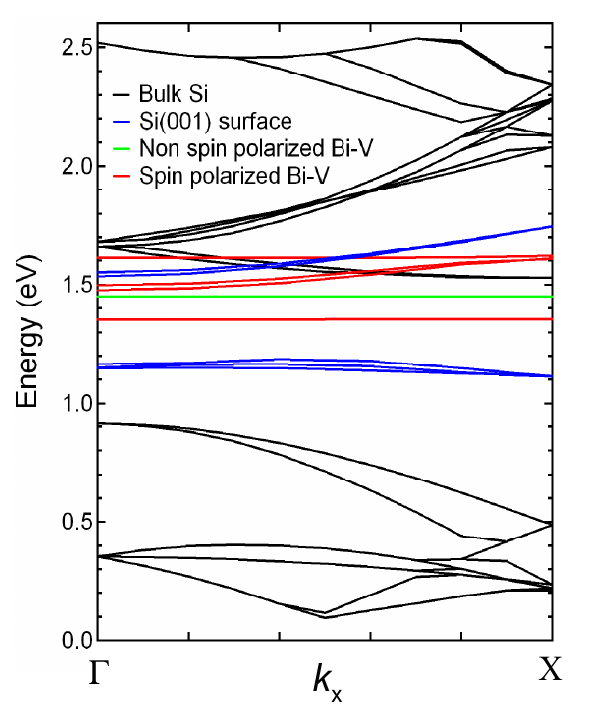}
  \caption{Band structure of bulk silicon,
    Si(001) and Bi-V feature along $\Gamma$-X (100) direction in the
    conventional cubic cell.}\label{fig:bands}
\end{figure}
 
We have discovered an unusual structure formed by the heavy dopant Bi
at Si(001) surfaces following ion-implantation and surface
cleaning. Fig. \ref{fig:bands}, where the band structures for bulk Si, Si(001)
and the Bi-V feature, with and without spin polarization, are plotted,
shows the essential physics. The reduction of the bandgap with
creation of the surface is clear, along with the introduction of
defect levels in the surface state gap. The on-site exchange, due to
the Coulomb interaction, splits majority from minority spins and opens
up a large gap for the Bi levels. Bi thus behaves quite unlike a bulk
dopant, or even a conventional light dopant (e.g., P) at the
surface. We understand the formation and nature of this structure, and
its absence for lighter dopants, as the natural consequence of the
much less significant bonding role of the outer $s$ orbital for a heavy
atom in column V of the periodic table. Beyond the new atomic
arrangement associated with the incorporation of Bi in the surface,
the electronic structure is unprecedented for a dopant at the surface
of Si: the Bi atom has a half-filled, localized orbital, which carries
an unpaired spin. Given the interest in ferromagnets based on
semiconductors, it is especially encouraging that the DFT results show
that superlattices of Bi-V defects favour ferromagnetic order, and
indicate exchange couplings as large as 0.03 eV between surface Bi
atoms. For the future, the calculations suggest that arrays of
closely-spaced Bi-V features would be fully polarized at accessible
temperatures. This would represent progress over a variety of
heterogeneous thin film systems\cite{Schmehl:2007ai,Jonker:2007tg} including Si as well as dilute
magnetic semiconductors such as Mn in GaAs\cite{Ohno:1998dp,Kitchen:2006hc} and Ge\cite{Park:2002sp}. The
half-filled orbitals which we have discovered provide an entr\'{e}e to Si(100) surface-based quantum many body physics, thus complementing previous work\cite{Santoro:1999ly} concerning adatom layers on Si(111), and could also find use in surface spin-based quantum information technology\cite{Stoneham:2003th,Kane:1998ij}.






\begin{acknowledgments}
This work was supported by the European Union, the Basic Technologies
Programme of the RCUK, a Wolfson-Royal Society Research Merit Award,
the EPSRC-funded programme COMPASSS, a
Royal Society URF and the Brazilian agencies CNPq and FAPESB.  We are
also very grateful to N. Curson, S. Schofield, C. Hirjibehedin,
\O . Fischer, J. P. de Souza and H. Boudinov for numerous helpful
discussions. 
\end{acknowledgments}


\appendix

\section{Comparison to heterodimers}
\label{sec:comp-heter}

\begin{figure}
  \centering
  \includegraphics[width=0.5\textwidth]{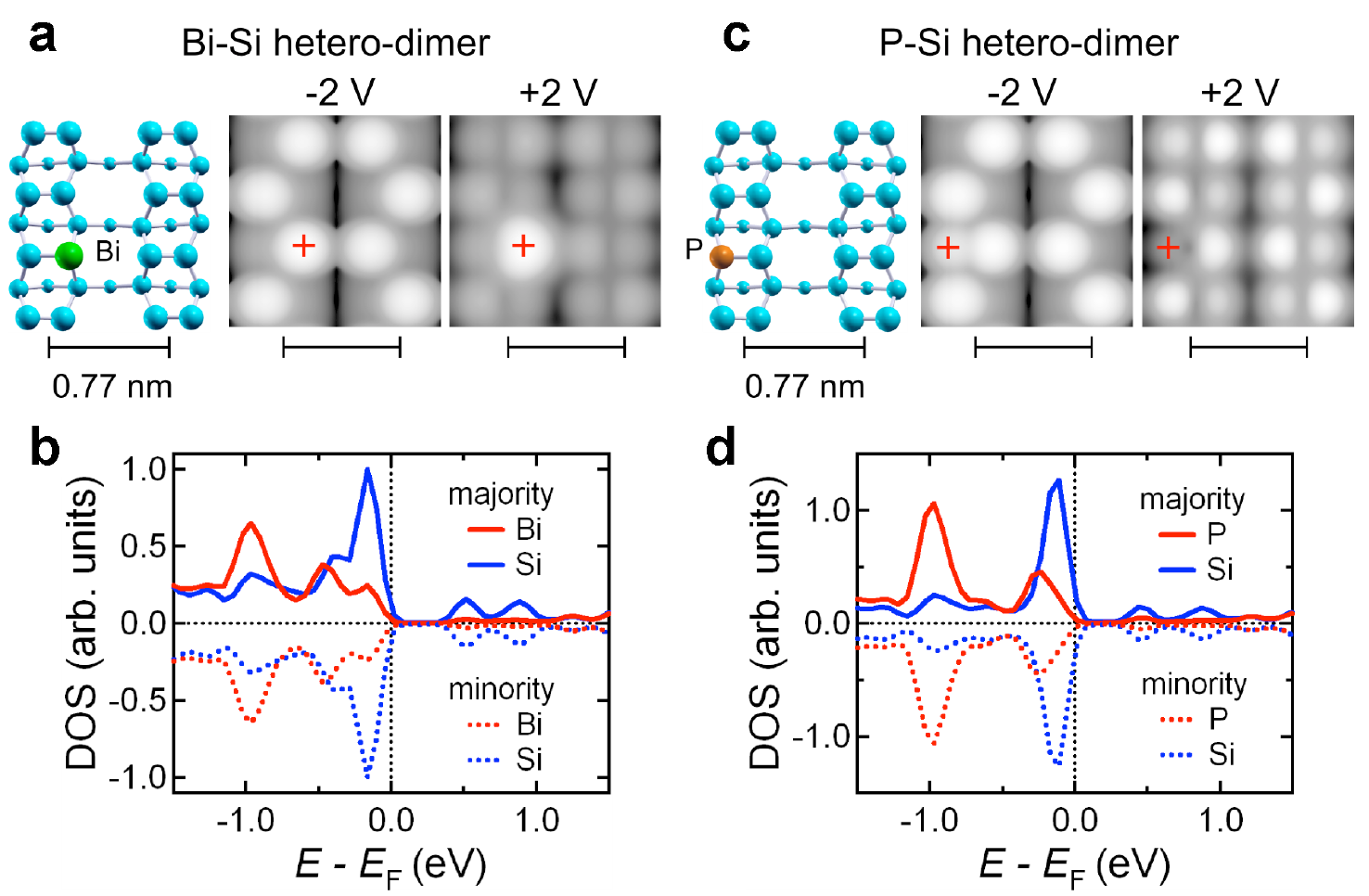}
  \caption{Simulated integrated DOS images and DOS of Bi-Si and P-Si
    hetero-dimers in Si(001), with the Bi and P atoms replacing the
    upper atom of a Si dimer in c(4$\times$ 2) reconstruction. (a)
    Left: Schematic structure of Bi-Si hetero-dimer.  The Bi atom has
    a neutral configuration. Right: Simulated integrated DOS images at
    Vs = -2 V, +2 V. Bi sites are shown by markers. (b) Spin-polarized
    DOS of Bi and Si atoms in the Bi-Si hetero-dimer structure shown
    in (a).  (c) Left: Schematic structure of P-Si
    hetero-dimer. Right: Simulated integrated DOS images at Vs = -2 V,
    +2 V, without the additional electron discussed in
    Ref.\onlinecite{Radny:2006fe}. The P sites are marked by crosses
    `+'. Both images are for isosurface values of 0.1 electron/nm$^3$.
    (d) Spin-polarized DOS of P and Si atoms in the P-Si hetero-dimer
    structure.  All integrated DOS images are for isosurface values of
    0.1 electron/nm$^3$.  }
\label{fig:Si3}
\end{figure}

As a further test of the correct assignment of the Bi-V structure to
the observed features, we have simulated STM and STS data for the
Bi-Si heterodimer, and compare it to the P-Si hetero dimer in
Fig.~\ref{fig:Si3}.  It is clear that this structure is not what is
observed in STM; it is also interesting to note that the STS
appearance for both P and Bi is extremely similar. 

\end{document}